\documentclass[
onecolumn,
superscriptaddress,
secnumarabic,
amssymb,
nobibnotes,
aps,
12pt,
longbibliography
]{revtex4-2}

\setcitestyle{numbers,square}
\usepackage[colorlinks=true,linkcolor=green, citecolor=green,allcolors=blue]{hyperref}

\usepackage{graphicx}
\usepackage{dcolumn}
\usepackage{bm}
\usepackage[utf8]{inputenc}
\usepackage[T1]{fontenc}
\usepackage{mathptmx}
\usepackage{etoolbox}
\UseRawInputEncoding

\makeatother
\usepackage{graphicx}
\usepackage{amsmath}
\usepackage{mathtools}
\usepackage{upgreek}
\usepackage{hhline}
\usepackage{here}
\usepackage{booktabs}
\usepackage{multirow}
\usepackage{array}   
\usepackage{tabularx}
\usepackage{textcomp,gensymb}
\usepackage[per-mode=symbol,range-phrase=--,range-units=single]{siunitx}
\usepackage{color}
\usepackage{nicefrac}
\usepackage{xcolor}
\usepackage{cleveref}
\usepackage[]{changes}
\usepackage{setspace}
\hypersetup{
    colorlinks=true,
    linkcolor=blue,
    filecolor=blue,      
    urlcolor=blue,
}

\newcommand{\ARCNL}{Advanced Research Center for Nanolithography (ARCNL),\\Science Park 106, 1098 XG Amsterdam, The Netherlands}
\newcommand{\VU}{LaserLab, Department of Physics and Astronomy, Vrije Universiteit Amsterdam,\\ De Boelelaan 1100, 1081 HZ Amsterdam, The Netherlands}

\newcommand{\We}{\mathrm{We}}

\newcommand{\tauc}{\tau_\mathrm{c}}
\newcommand{\taup}{\tau_\mathrm{p}}

\newcommand{\el}{E_\mathrm{L}}

\begin{document}

\renewcommand{\thesection}{\Roman{section}}
\renewcommand{\thesubsection}{\Alph{subsection}}
\renewcommand{\thesubsubsection}{\arabic{subsubsection}}

\singlespacing 
\title{Ambient-pressure effects on shock emission and liquid-shell expansion following optical breakdown inside water droplets}

\author{M. Brugge}
\thanks{These authors contributed equally to this work}
\affiliation{\ARCNL}

\author{S. van Belois}
\thanks{These authors contributed equally to this work}
\affiliation{\ARCNL}

\author{S.S. Ade}
\affiliation{\ARCNL}

\author{O.O. Versolato}
\email{versolato@arcnl.nl}
\affiliation{\ARCNL}
\affiliation{\VU}

\author{M. Kharbedia}
\email{m.kharbedia@gmail.com}
\affiliation{\ARCNL}

\date{\today}

\begin{abstract}
We experimentally investigate the deformation and breakup of micrometer-sized water droplets following nanosecond laser-induced optical breakdown in the finite-beam regime $D_0/D_\mathrm{b}\simeq2$ and over a range of ambient pressures. Simultaneous measurements of the liquid deformation and emitted shock wave enable the gas and liquid phase dynamics to be distinguished quantitatively. Under fixed external focusing conditions, the curved droplet interface focuses and internally reflects the incident light, leading to transient plasma formation within the liquid and generating the pressure load that drives droplet deformation. At low laser energies ($\el<1\,\mathrm{mJ}$), plasma formation localized near the rear droplet-air interface produces an interfacial bubble that expands into a thin liquid film susceptible to hydrodynamic instabilities and ultimately collapses under the action of surface tension. The internal pressure is partially released into the surrounding air, generating a propagating shock wave. At higher energies ($\el>1\,\mathrm{mJ}$), the droplet is ablated along the laser axis, resulting in the explosive formation of a liquid shell and multiple shock fronts, both exhibiting supersonic expansion velocities. While the shock wave dynamics are well described within the framework of strong blast wave theory, we show that the shell expansion can be understood as being driven by the pressure load generated during plasma expansion inside the droplet. Consequently, the droplet expands as a liquid shell that subsequently follows ballistic motion and ultimately undergoes complete atomization. Although changing the surrounding gas density substantially alters the propagation of the emitted shock wave, the droplet morphology and expansion dynamics remain nearly unchanged, given the internal load of the plasma-driven pressure. An energy partition analysis indicates that up to $\sim70\%$ of the incident laser energy is divided between the shock wave and the expanding liquid shell, regardless of the ambient pressure, with the remaining energy distributed among other channels, including plasma radiation, liquid heating, and laser light transmission.
\end{abstract}\def\keywordstitle{Keywords}

\maketitle

\onecolumngrid

\section{Introduction}\label{Introduction}

\begin{figure}
    \centering
    \includegraphics[width=0.6\textwidth]{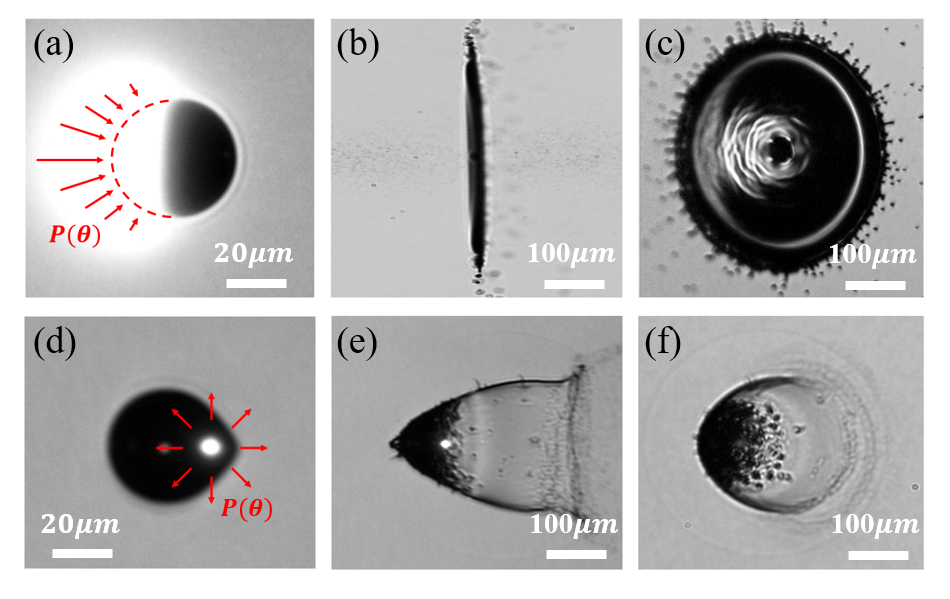}  
    \caption{Examples of two laser-droplet interaction regimes, depending on the optical transmittance of the liquid. (a) Deformation of a micron-sized liquid tin droplet following nanosecond laser impact (incident from the left), leading to the formation of a transient hot plasma. The recoil pressure, $P(\theta)$ (with $\theta$ the polar angle), generated by the expanding plasma drives the transformation of the initially spherical droplet into a thin, radially expanding sheet, shown in side and front views in (b) and (c), respectively. (d) Laser impact on a similarly sized water droplet (also incident from the left), for which the transparent medium enables light focusing within the droplet. This leads to optical breakdown inside the liquid and the generation of spherically expanding, plasma-driven pressure waves. The late-time dynamics differ significantly from the tin case, as illustrated by the side and front views in (e) and (f).}
    \label{fig:introduction}
\end{figure}

Laser-droplet interactions underpin numerous applications, including extreme-ultraviolet (EUV) light generation for industrial nanolithography\,\cite{versolato2019physics,fu2019euv_review, versolato2022microdroplet}, hydrogen-droplet targets for inertial-fusion experiments\,\cite{nordhage2005_hydrogen_drop_fusion, fraga2012_hydrogen_drop_laser_plasma}, and laser-produced plasmas for particle-acceleration research\,\cite{karsch2003drop_particle_laser, ter2005fusion_laser_drop}. Additional applications include laser-induced breakdown spectroscopy for the chemical analysis of droplets\,\cite{contreras2018_LIB_drop_chemical, palleschi2025_LIB_review_drop,Lokini_2024_LIB_heptane_droplets}, inkjet printing and additive manufacturing for nanoparticle sintering\,\cite{Lohse_inkjet_2022,Liu2023_Inkjet}, printed electronics\,\cite{inui_laser-induced_2015,nguyen2026transfer}, needle-free injection\,\cite{kale2014needle, barolet2018current_needle_free}, spray-combustion diagnostics \,\cite{Enomoto2017,fragmentationofhydrocarbonfuel2024}, laser-driven propulsion on wetting and nonwetting surfaces \,\cite{Han_2025_laser_pulling_drop_superhydrophobic,Ahn2013}, and laser-induced cavitation followed by jetting \,\cite{gonzalez_avila_fragmentation_2016,Rao2022,Reuter2023}. For opaque droplets, the laser energy is deposited within a thin surface layer, producing a rapidly expanding laser-absorbing plasma whose recoil pressure transforms the droplet into a radially expanding liquid sheet\,\cite{klein_drop_2020,kurilovich2018power, hernandez2022early,Kharbedia2026_Rim}. The resulting deformation dynamics are well understood and are governed primarily by the recoil pressure of the expanding plasma\,\cite{Kharbedia2026_LWO, Kharbedia2026_Rim,francca2025laser,liu_2023_mass}. For transparent liquids, by contrast, transmission through the droplet alters the optical-field distribution and makes plasma formation strongly dependent on the focusing geometry (see Fig.\,\ref{fig:introduction}). 

Laser-induced processes in transparent micrometer-sized droplets were investigated extensively in early optical-breakdown studies \,\cite{alexander1987explosive_vap_drop, eickmans1987drop_explosion_laser_spectra, chang1988laser_transparent_drop, hsieh1987_plasma_front_laser_trans_drop, zhang1987explosive_vap_larg_drop_laser, kafalas1973fog_vaporization_10um_drop, wood1988trans_drop_laser_water_ethanol}. When the droplet dimensions exceed the optical wavelength, the curved liquid-gas interface redistributes the incident field and produces pronounced intensity maxima within the liquid and in the shadow-side near field. These internal and external maxima were established through Lorenz-Mie calculations and spatially resolved optical measurements \,\cite{benincasa1987spatial,chang1988laser_transparent_drop}. Plasma formation was subsequently observed both within the liquid and in the near-field gas. The breakdown-initiation site depends on the droplet size and refractive index, the liquid and gas breakdown thresholds, the surrounding gas and pressure, and the laser wavelength and pulse duration\,\cite{chang1988laser_transparent_drop, chylek1987_LIB_drop_gases,chylek1990pressure}. Two disruption mechanisms must be distinguished in this literature. At strongly absorbed wavelengths, rapid heating can produce explosive vaporization, whereas at weakly absorbed wavelengths sufficiently high local intensities initiate optical breakdown and plasma-driven fragmentation\,\cite{kafalas1973fog_vaporization_10um_drop, alexander1987explosive_vap_drop,zhang1987explosive_vap_larg_drop_laser}. Previous studies also established plasma-front propagation through and outside transparent droplets \,\cite{hsieh1987_plasma_front_laser_trans_drop, eickmans1987drop_explosion_laser_spectra, chang1988laser_transparent_drop}. For the present finite-beam geometry, however, the relative contributions of the transmitted and internally reflected branches to the pre-breakdown intensity distribution require separate quantification. Moreover, although ambient pressure is known to influence the breakdown threshold and initiation site\,\cite{chylek1990pressure}, its effect on the subsequent droplet deformation, fragmentation, and emitted shock-wave dynamics has received comparatively little quantitative attention. In particular, a unified comparison of the liquid-shell and shock-front trajectories across pulse energy and ambient pressure is still lacking. A major branch of recent hydrodynamic studies has considered millimeter-scale levitated droplets irradiated by tightly focused micrometer-scale beams, such that $D_0/D_b\gg1$ \,\cite{gonzalez_avila_fragmentation_2016,Singh_2026}. Because the focal region is much smaller than the droplet radius, the illuminated interface can be treated locally as planar, and the breakdown location is prescribed primarily by the externally imposed focus. It has been shown that cavitation-driven deformation and fragmentation depend sensitively on the bubble location \,\cite{gonzalez_avila_fragmentation_2016}. More recently, it has been shown that breakup morphology and propulsion direction can be selected through the combined control of pulse energy and axial droplet placement relative to the external laser focus \,\cite{Rao2022, Singh_2026}.

Here, we investigate nanosecond infrared irradiation of micrometer-sized water droplets in the regime $D_0/D_b\simeq2$ over a range of ambient pressures. The ambient pressure is used as a dynamical control parameter to distinguish the influence of the surrounding gas on shock propagation from its influence on the deformation and fragmentation of the liquid. The external beam geometry and droplet placement are kept fixed, leaving the pulse energy as the only varied optical parameter at each ambient pressure. Slightly above the breakdown threshold, plasma formation near the rear interface produces an interfacial bubble that expands into a thin liquid film and subsequently retracts under capillary forces. Increasing the pulse energy extends the plasma along the droplet axis, producing stronger ablation, multiple shock fronts, supersonic liquid-shell expansion, and ultimately complete fragmentation. Simultaneous measurements of the external shock and liquid-shell trajectories at different ambient pressures enable their dynamics to be separated quantitatively. Within the strong-shock and approximately spherical regime, the external shock follows a density-dependent Taylor-Sedov-type scaling\,\cite{taylor1950formation, Ramsden1964_radiative_blast,ruban_fluid_dynamics_I}, whereas the liquid shell exhibits ballistic expansion that is nearly independent of ambient pressure. Rescaling the shock and shell trajectories according to their respective energy and density dependencies collapses measurements obtained across the explored pulse energies and ambient pressures. 
Although pressure-dependent optical-breakdown thresholds in air have been reported previously\,\cite{chylek1990pressure}, the pressure dependence of the coupled post-breakdown shock and liquid-shell dynamics has not been characterized quantitatively. In the present study, changing the ambient gas density alters the propagation of the external shock while leaving the liquid-shell morphology and expansion rate nearly unchanged. This separation demonstrates that the dominant deforming impulse originates inside the liquid, even though the subsequent shock propagation is governed by the surrounding gas. Finally, the estimated shock-wave and liquid kinetic energies together account for an upper value of approximately $70\%$ of the incident laser energy, indicating substantial coupling of the optical energy to mechanically resolved gas and liquid motion.

This manuscript is organized as follows. In Sec.\,\ref{sec:experimental} we describe the experimental setup. In Sec.\,\ref{sec:results}\,\ref{subsec:optical_breakdown}, we characterize plasma formation using time-resolved imaging and a two-dimensional ray-tracing model. Section\,\ref{sec:results}\,\ref{subsec:phase_diagram} presents the observed deformation regimes, followed by an analysis of thin-film dynamics in Sec.\,\ref{sec:results}\,\ref{subsec:thin_film_deformation}. In Sec.\,\ref{sec:results}\,\ref{subsec:droplet_shock_dynamics}, we quantify the coupled droplet and shock-wave dynamics, while Sec.\,\ref{sec:results}\,\ref{subsec:energy_partition} discusses the associated mechanical-energy channels and the broader implications of the results.

\section{Experiment}\label{sec:experimental}

\begin{figure}
    \centering
    \includegraphics[width=0.8\textwidth]{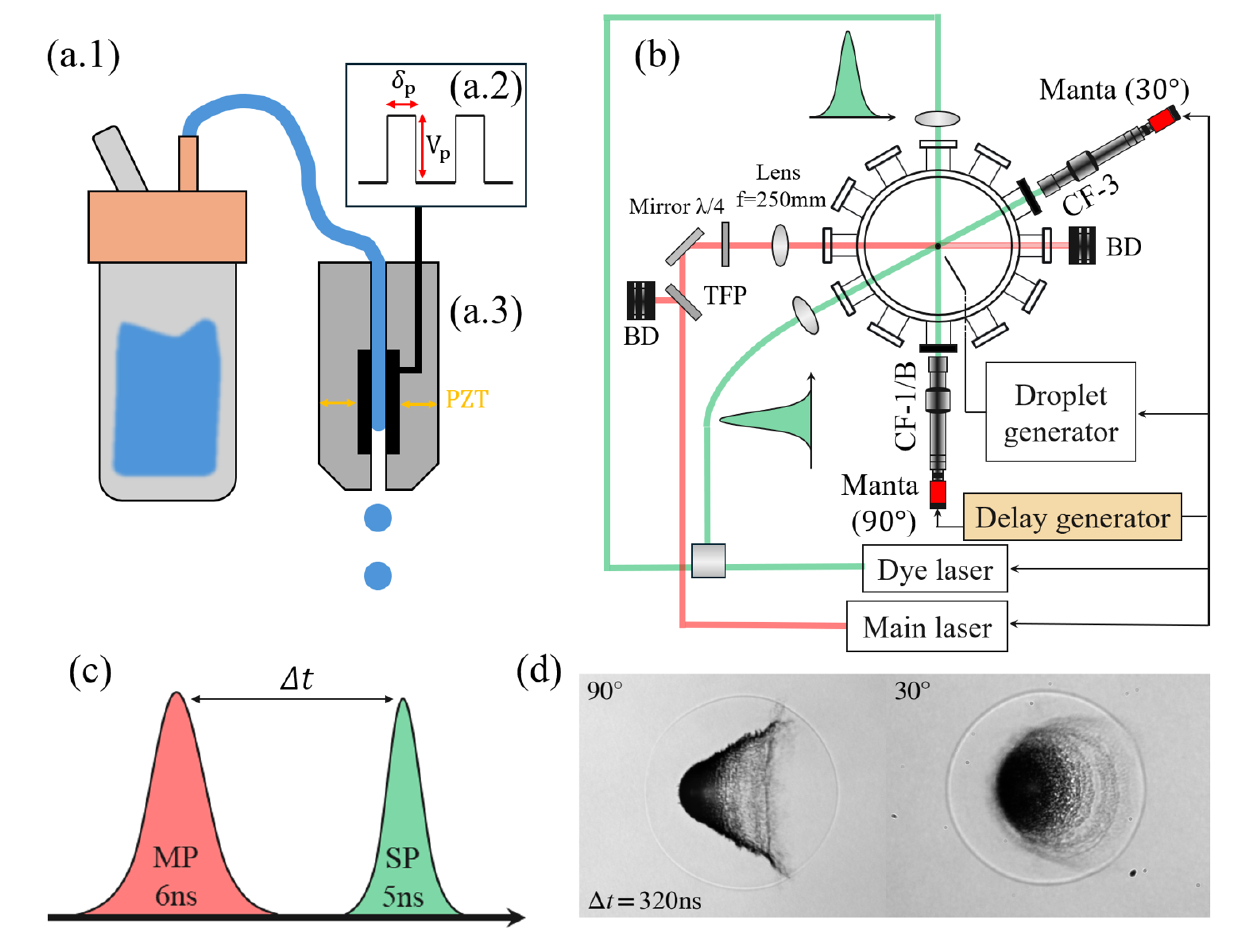}  
    \caption{Experimental setup used in this study. (a) Conceptual sketch of the droplet-on-demand generator. (a.1) A vial filled with water is connected via a capillary tube to a nozzle equipped with a piezoelectric crystal (PZT). (a.2) The PZT is driven by a square pulse with amplitude $V_\mathrm{p}$ and pulse duration $\delta_\mathrm{p}$. (a.3) The resulting pressure generated in the nozzle produces droplets on demand with a $D_0=90\,\mu\mathrm{m}$ diameter. (b) Top-view illustration of the measurement setup. The droplet is injected into a chamber and irradiated by the main laser pulse with temporal duration $\tau_\mathrm{p}=6\,\mathrm{ns}$. Shadowgraph images are recorded using CCD cameras after illuminating the droplet with spatially non-coherent dye laser pulses of duration $\tau_\mathrm{s}=5\,\mathrm{ns}$ from two different angles, providing a side view ($90^\circ$) and a front view ($30^\circ$). The timing of the system, including droplet generation, main pulse impact, shadowgraphy pulse emission, and image acquisition, is controlled by a delay generator with internal triggering. (c) The time evolution of the droplet after laser impact is captured by varying the delay of the shadowgraphy pulse (SP) with respect to the main pulse (MP) in the range $\Delta t=2\,\mathrm{ns}$-$2\,\mu\mathrm{s}$. (d) Example droplet profiles at $\Delta t=320\,\mathrm{ns}$ from both viewing angles.}
    \label{fig:experimental_setup}
\end{figure}

Figure\,\ref{fig:experimental_setup} illustrates the experimental setup used during the experiments. We produce water droplets with a commercial on demand micro-droplet generator (Microdrop Technologies GmbH). The liquid contained in a vial is connected to a nozzle provided with a piezo that produces a pressure difference driven by external voltage, as shown in Fig.\,\ref{fig:experimental_setup}(a), releasing droplets with diameters $D_0=50-100\,\mu\mathrm{m}$, depending on the voltage peak value $V_\mathrm{p}$ and pulse duration $\delta_\mathrm{p}$. For our experiments, a fixed droplet diameter with $D_0=90\,\mu\mathrm{m}$ is produced with $V_\mathrm{p}=100\,\mathrm{mV}$ and $\delta_\mathrm{p}=45\,\mu\mathrm{s}$. 
These droplets are injected into a vacuum chamber, where the air pressure can be reduced from $1\,\mathrm{bar}$ to $0.1\,\mathrm{bar}$. Each droplet is irradiated by a circularly polarized Q-switched Nd:YAG laser pulse, with pulse length $\taup\sim 6\,\mathrm{ns}$, and wavelength $\lambda_\mathrm{L}=1064\,\mathrm{nm}$. 
The laser energy is controlled using a half-wave plate and a thin-film polariser, providing a peak energy range of $0.4-200\,\mathrm{mJ}$. 
The beam, with an average diameter of $\sim6\,\mathrm{mm}$, is focused at the droplet as a Gaussian spot with a plano-convex focal lens ($f=250\,\mathrm{mm}$), resulting in $\sim45\,\mu\mathrm{m}$ diameter at full width at half maximum (FWHM). 
We illuminate the droplet with a temporally and spatially incoherent green light with wavelength $564\,\mathrm{nm}$ and temporal width $\sim5\,\mathrm{ns}$, and record the resulting shadow with CCD cameras (Manta G145-B, AVT) provided with long-distance microscopes (CF-1/B, CF-3; Edmund Optics). The optical resolution achieved with this setup is approximately $\sim5\,\mu\mathrm{m}$. To visualize the laser-droplet interaction, we use a stroboscopic system, recording the imaged shadow of the droplet from the orthogonal angle of the main laser beam (that is, at $90^\circ$) and nearly front view ($30^\circ$), as illustrated in Fig.\,\ref{fig:experimental_setup}(b). 
To precisely control the timing of laser interaction with the droplet and image recording, we use a delay generator (Stanford Research DG645). The entire electronic system (that is, droplet generator, main pulse, shadowgraphy pulse, and cameras) is triggered with a $10\,\mathrm{Hz}$ repetition rate, which results in ten laser-droplet impact events per second.
The temporal evolution of the droplet following the laser impact is recorded by delaying the shadowgraphy pulse (SP) with respect to the main laser pulse (MP; hereafter, referred to as simply the laser pulse), with separation times between these two pulses ranging $2\,\mathrm{ns}-2\,\mu\mathrm{s}$, with times steps within $2-50\,\mathrm{ns}$. For each time step, we record several frames for a better statistical analysis (typically between $10-20$ frames per step). 
Before any measurement, we carry out an initial optimization of the laser to droplet alignment by observing the tilted angle of the film at a given energy. We found negligible horizontal and vertical drift in droplet release. The vertical drift is an average $<2\,\mu\mathrm{m}$ per droplet release, without significantly affecting the reproducibility of the laser-droplet interaction. Simultaneously, the laser beam has an average y-z drift $\pm5\,\mu\mathrm{m}$ over a run during several minutes. The typical ratio of the droplet to beam size is $\alpha=D_0/D_\mathrm{b}\sim2$, with $D_0\sim90\,\mu\mathrm{m}$ and $D_\mathrm{b}\sim45\,\mu\mathrm{m}$ (corresponding to FWHM). With $\alpha\sim2$ and an approximately collimated beam, the droplet surface acts as a strong lens that focuses the light inside. Although the incident intensity remains below the bulk-water breakdown threshold, optical focusing within the droplet raises the local intensity sufficiently to produce breakdown under the experimentally realized alignment conditions. The measured alignment drift does not prevent plasma formation, although they may contribute to variations in the plasma morphology and subsequent droplet deformation.

\section{Results and discussion}\label{sec:results}

\begin{table*}[t]
    \centering
    \setlength{\tabcolsep}{5pt}
    \caption{Relevant experimental parameters: Droplet diameter $D_0$, capillary timescale $\tauc$, acoustic timescale $\tau_\mathrm{a}$, laser pulse energy $\el$, radial expansion rate for droplet $\dot{R}_\mathrm{d}$, the corresponding Weber number $\We$, Mach number of the droplet expansion $\mathrm{M}_\mathrm{d}$, shock wave radial expansion rate $\dot{R}_\mathrm{sh}$, the corresponding shock Mach number $\mathrm{M}_\mathrm{sh}$, and the ambient pressure $p_0$.} 
    \renewcommand{\arraystretch}{1.5}
    \begin{tabular}{ccccc ccccc}
        \hline
        $D_0$ ($\mu\mathrm{m}$) & $\tauc$ ($\mu\mathrm{s}$) & $\tau_\mathrm{a}$ ($\mathrm{ns}$) & $E_\mathrm{L}$ (mJ) & $\dot{R}_\mathrm{d}$ (m/s) & $\We$ & $\mathrm{M}_\mathrm{d}$ & $\dot{R}_\mathrm{sh}$ (m/s) & $\mathrm{M}_\mathrm{sh}$ & $p_0$ (bar) \\
        \hline
        90 & 41.1 & 30 & 0.4--8.0 & 100-4200 & $>2000$ & 0.4-12 & 400-4500 & 1-13 & 0.1-1\\
        \hline
    \end{tabular}
    \label{table:exp_data}
\end{table*}

A quantitative description of the droplet and shock wave dynamics requires the introduction of several dimensionless parameters. For low-energy laser pulses, the balance between inertial and capillary forces is captured by the Weber number, $\We$, while the characteristic timescale is the capillary time, $\tauc$. At higher energies, where droplet expansion becomes supersonic and shock wave effects dominate, we introduce the Mach number $\mathrm{M}$, defined as the ratio between a characteristic velocity and the speed of sound in air. Furthermore, to characterize the relevant timescales of pressure-wave propagation within the droplet, we define an acoustic timescale, $\tau_\mathrm{a}$. The mentioned parameters are defined as:

\begin{equation}\label{eq:dimensionless}
    \mathrm{We}=\frac{\rho\dot{R}^2D_0}{\sigma} \,;\,\,\,\, \tauc=\sqrt{\frac{\rho D_0^3}{6\sigma}}\,;\,\,\,\, \mathrm{M}=\frac{\dot{R}}{c}\,;\,\,\,\,\tau_\mathrm{a}=\frac{R_0}{c_\mathrm{w}}.
\end{equation}

Here, $\rho=1000\,\mathrm{kg/m^3}$ and $\sigma=0.072\,\mathrm{N/m}$ are the density and surface tension of water at room temperature, respectively. The characteristic expansion rate is defined as $\dot{R}$ (subscripts will be used to distinguish between droplet and shock waves), and the speed of sound in air $c\sim340\,\mathrm{m/s}$, and in water $c_\mathrm{w}\sim1480\,\mathrm{m/s}$. Typical experimental parameters explored in this study are listed in Table\,\ref{table:exp_data}.

\subsection{Laser-droplet interaction: optical breakdown}\label{subsec:optical_breakdown}

\begin{figure}
    \centering
    \includegraphics[width=1\textwidth]{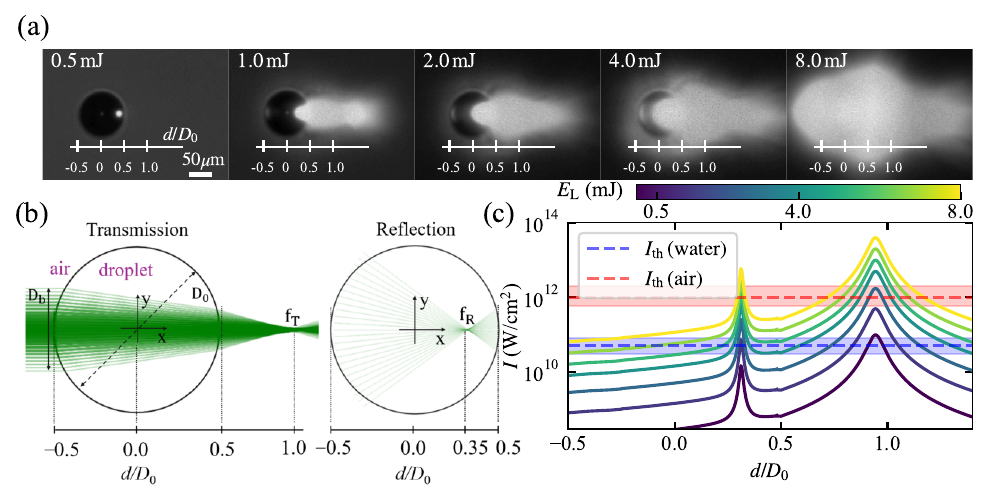}  
    \caption{Laser-induced optical breakdown in water droplets. (a) Evolution of the plasma front with increasing laser energy. An additional horizontal axis indicates the dimensionless axial coordinate $d/D_0$. (b) Schematic representation of the ray-tracing model (see Appendix\,\ref{app:ray-tracer}). Transmitted (left) and internally reflected (right) rays are shown separately, with transmission and reflection intensity ratios of $I_\mathrm{t}/I_0=0.98$ and $I_\mathrm{r}/I_0=0.02$, respectively. Note the second transmission into the air corresponds to $I'_\mathrm{t}=0.98I_\mathrm{t}$. The Gaussian beam profile with diameter $D_\mathrm{b}$ produces two focal regions: an internal focus $\mathrm{f_R}$ generated by reflection at $d/D_0\sim0.35$, and a second focus $\mathrm{f_T}$ in air at $d/D_0\sim1$. (c) Calculated intensity distribution along the optical axis. Inside the droplet ($-0.5<d/D_0<0.5$), the total intensity is given by the sum of the transmitted and reflected contributions, $I_\mathrm{t}+I_\mathrm{r}$. Two pronounced intensity maxima are observed at $d/D_0\sim0.35$, corresponding to the internally reflected focus $f_\mathrm{R}$, and at $d/D_0\sim1$, corresponding to the focus formed in air, $f_\mathrm{T}$. The curves are shown for laser pulse energies in the range $E_\mathrm{L}=0.2-8.0\,\mathrm{mJ}$. Dashed lines indicate previously reported optical breakdown thresholds for water (blue) and air (red) under nanosecond laser irradiation. Shaded bands represent literature-derived optical breakdown threshold ranges for water and air, centered on the mean values obtained from multiple published studies\,\cite{vogel_shock_1996,vogel_mechanisms_2005}.}
    \label{fig:optical_breakdown}
\end{figure}

Figure\,\ref{fig:optical_breakdown} shows a sequence of images recorded at the instant of laser impact, $t=0\,\mathrm{ns}$, for different laser pulse energies. The axial coordinate along the laser propagation direction is normalized by the droplet diameter, $d/D_0$, where $d/D_0=0$ corresponds to the droplet center and $d/D_0=\pm0.5$ to the liquid-air interfaces. The lowest pulse energy at which plasma formation is observed is approximately $0.5\pm0.1\,\mathrm{mJ}$, with the plasma localized near the rear side of the droplet. Previous studies have reported that the intensity threshold for nanosecond laser-driven optical breakdown in water is of the order of $\sim10^{11}\,\mathrm{W/cm^2}$\,\cite{vogel_shock_1996, vogel_mechanisms_2005}. Under the present conditions, with the laser energy $\el=0.5\,\mathrm{mJ}$, pulse duration $\taup\sim6\,\mathrm{ns}$, and the Gaussian spot diameter $D_b\sim45\,\mu\mathrm{m}$ at FWHM, the resulting peak intensity can be estimated as: 

\begin{equation}\label{eq:laser_intensity}
    I_0\approx\frac{2P_\mathrm{p}}{\pi\omega_0^2}=\frac{4\el\ln{2}}{\pi\taup D_b^2}\approx3.6\times10^{9}\,\mathrm{W/cm^2},
\end{equation}
where $\omega_0=D_b/\sqrt{2\ln{2}}$ is the $1/\mathrm{e}^2$ beam radius, and $P_\mathrm{p}\approx \el/\taup$ is the peak power of the pulse. This value is approximately 30 times lower than the reported threshold and highlights the crucial role of internal focusing in plasma formation inside the droplet. To elucidate this focusing mechanism, Fig.\,\ref{fig:optical_breakdown}(b) presents a two-dimensional ray-tracing simulation. Given the droplet size and laser wavelength, $\lambda/D_0\ll1$, the geometrical-optics approximation is well justified\,\cite{quick2024_ball_lens}. The energy carried by each ray is weighted according to the Gaussian intensity profile of the incident beam. A detailed description of the model is provided in Appendix\,\ref{app:ray-tracer}. Here, we consider parameters representative of the experiments, with $\alpha=D_0/D_b=2$. At normal incidence, approximately $98\%$ of the optical power is transmitted at each water-air interface, whereas approximately $2\%$ is reflected\,\cite{born2013principles_optics}. In this simplified calculation, the transmitted and internally reflected branches are initially weighted using normal-incidence Fresnel coefficients\,\cite{hecht_optics_2002}. The quantitative model presented in Appendix\,\ref{app:ray-tracer} subsequently accounts for the dependence of the Fresnel coefficients on the incidence angle, including total internal reflection where applicable. 

It is important to note that the ray-tracing calculation describes the linear optical field prior to breakdown. Once plasma is formed, absorption, scattering, plasma shielding, and refractive-index changes can substantially modify the remaining part of the laser pulse \,\cite{chylek1987_LIB_drop_gases,chylek1990pressure, chang1988laser_transparent_drop,zhang1987explosive_vap_larg_drop_laser}. Consequently, the calculated transmission and reflection coefficients should not be interpreted as a measurement of the post-breakdown energy partition. Because the transmitted pulse energy is not measured in the present experiments, the absorbed fraction cannot be determined quantitatively. For simplicity, only the first internal reflection is considered. The model predicts an internal intensity maximum near $d/D_0\simeq0.35$ and an external shadow-side focus near $d/D_0\simeq1$. The location of the internal maximum agrees closely with the internal-field structure reported in earlier Lorenz-Mie calculations and optical measurements of transparent droplets\,\cite{benincasa1987spatial,chang1988laser_transparent_drop}. The external focus near $d/D_0\simeq1$ is likewise consistent with the previously reported shadow-side near-field intensity maximum\,\cite{benincasa1987spatial,chang1988laser_transparent_drop,chylek1990pressure}. These studies identified plasma initiation either within the liquid or in the near-field gas, depending on droplet size, surrounding gas, and laser conditions. Subsequent time-resolved measurements further showed axial plasma-front propagation through the droplet and the formation of plasma plumes on both the shadow and laser-facing sides. The present ray tracing analysis is therefore not intended to establish microlensing-induced breakdown as a new mechanism. Rather, it validates the optical initial condition for the present geometry, $D_0/D_b\simeq2$, and separates the transmitted and internally reflected branches responsible for the two observed focal regions. The location of the transmitted focal region can be estimated independently using the paraxial ball-lens relations\,\cite{quick2024_ball_lens}:

\begin{align}\label{eq:ball_lens}
    \mathrm{EFL}=\frac{n_\mathrm{w}D_0}{4(n_\mathrm{w}-1)},
\end{align}
where $\mathrm{EFL}$ denotes the effective focal length. This quantity defines the focal-point location measured from the droplet center ($\mathrm{EFL}$). For water, with $n_\mathrm{w}\approx1.33$ and $D_0\approx90\,\mu\mathrm{m}$, one obtains $\mathrm{EFL}\approx90\,\mu\mathrm{m}$, corresponding to $d/D_0\approx1$. This estimate agrees closely with the focusing location obtained from the ray-tracing model. The position of the internally reflected focus can be estimated by treating the rear interface as a concave spherical mirror with focal length $f_\mathrm{m}\simeq R_0/2$. The external ball-lens focus acts as a virtual object for this interface. Applying the spherical-mirror equation places the reflected image approximately $15\,\mu\mathrm{m}$ inside the rear surface, corresponding to $d/D_0\simeq0.33$. This value agrees closely with the location $\mathrm{f_R}$ obtained from the ray tracing model (see Fig.\,\ref{fig:optical_breakdown}).
 
While geometrical optics predicts that refracted rays converge to a point within the droplet, this picture neglects the finite wavelength of the laser radiation. In practice, diffraction imposes a lower bound on the achievable focal size under ideal paraxial Gaussian focusing\,\cite{quick2024_ball_lens}:
\begin{equation}\label{eq:diffraction_limit}
    w_\mathrm{diff}\approx\frac{\lambda_\mathrm{L}}{\pi\mathrm{NA}},
\end{equation}
where $\mathrm{NA}$ denotes the numerical aperture. For paraxial rays, the numerical aperture can be approximated as:
\begin{equation}\label{eq:numerical_aperture}
    \mathrm{NA}=\frac{1}{\sqrt{1+4\left(\frac{\mathrm{EFL}}{D_\mathrm{b}}\right)^2}},
\end{equation}
which yields $\mathrm{NA}\approx0.24$ for the present geometry. Consequently, the diffraction-limited focal size is $w_\mathrm{diff}\sim1.4\,\mu\mathrm m$. For consistency, we define all beam widths in the following as equivalent Gaussian $1/e^2$ intensity radii. 
The ray-tracing simulations predict minimum beam widths of approximately $w_\mathrm{geo}\simeq5\,\mu\mathrm{m}$. For consistency, we define the effective beam width as $w_\mathrm{eff}^2(x) = w_\mathrm{geo}^2(x)+w_\mathrm{diff}^2$. This gives an effective beam width at the focal point $w\simeq5.7\,\mu\mathrm{m}$. Therefore, these simulations provide a reasonable first-order estimate of the intensity enhancement (see Appendix\,\ref{app:ray-tracer} for details). The spherical geometry introduces significant aberrations, producing an elongated focal region rather than a point focus. The simulation predicts that the reflected branch raises the local intensity near $d/D_0\simeq0.35$ to values comparable to or exceeding reported nanosecond breakdown thresholds in water\,\cite{Vogel_LIP_thresholds}. 
Finally, the thermal diffusion timescale across the droplet, $\tau_k=D_0^2/4\alpha_\mathrm{k}\sim14\,\mathrm{ms}$, with $\alpha_\mathrm{k}=1.43\times10^{-7}\,\mathrm{m^2/s}$, is much longer than both the laser pulse duration and the characteristic plasma formation timescale, $\tau_h\sim\taup\sim6\,\mathrm{ns}$. Thermal diffusion therefore cannot redistribute energy over the droplet during the initial interaction. This does not imply that the localized plasma heating is negligible; rather, the deposited energy remains confined near the breakdown region on the nanosecond timescale.  

As the laser energy increases, a larger fraction of the liquid is ablated, resulting in an expansion of the plasma volume and, consequently, a stronger pressure load within the droplet [Fig.\,\ref{fig:optical_breakdown}(a)]. Notably, the plasma front extends in both directions along the laser-propagation axis. At $\el\gtrsim1\,\mathrm{mJ}$, a bright forward plasma plume becomes visible in the gas beyond the shadow-side interface, consistent with the increasing contribution of the transmitted optical field. At higher energies, two mechanisms may contribute to the axial growth of the plasma: direct breakdown induced by the transmitted field and avalanche ionization seeded by the plasma initially generated near the internally reflected focus. At $\el=8.0\,\mathrm{mJ}$, the plasma extends along the full droplet axis and emerges beyond the laser-facing interface into air ($d/D_0<-0.5$). This occurs even though the calculated breakdown intensity in this region remains below the breakdown threshold of undisturbed air. The observation is therefore consistent with breakdown-front propagation seeded by the existing plasma, rather than with direct breakdown of the surrounding gas by the incident beam alone. In the present geometry, the spatial extent of the plasma is tuned under fixed external focusing conditions solely by varying the pulse energy. The earliest resolved plasma emission remains localized inside the droplet near the reflected focus. Together with the optical model and the ambient-pressure measurements, this supports the interpretation that the dominant pressure load responsible for droplet deformation originates inside the liquid (see Sec.\,\ref{sec:results}\,\ref{subsec:droplet_shock_dynamics}).
 
\subsection{Phase diagram}\label{subsec:phase_diagram}

\begin{figure}
    \centering
    \includegraphics[width=1\textwidth]{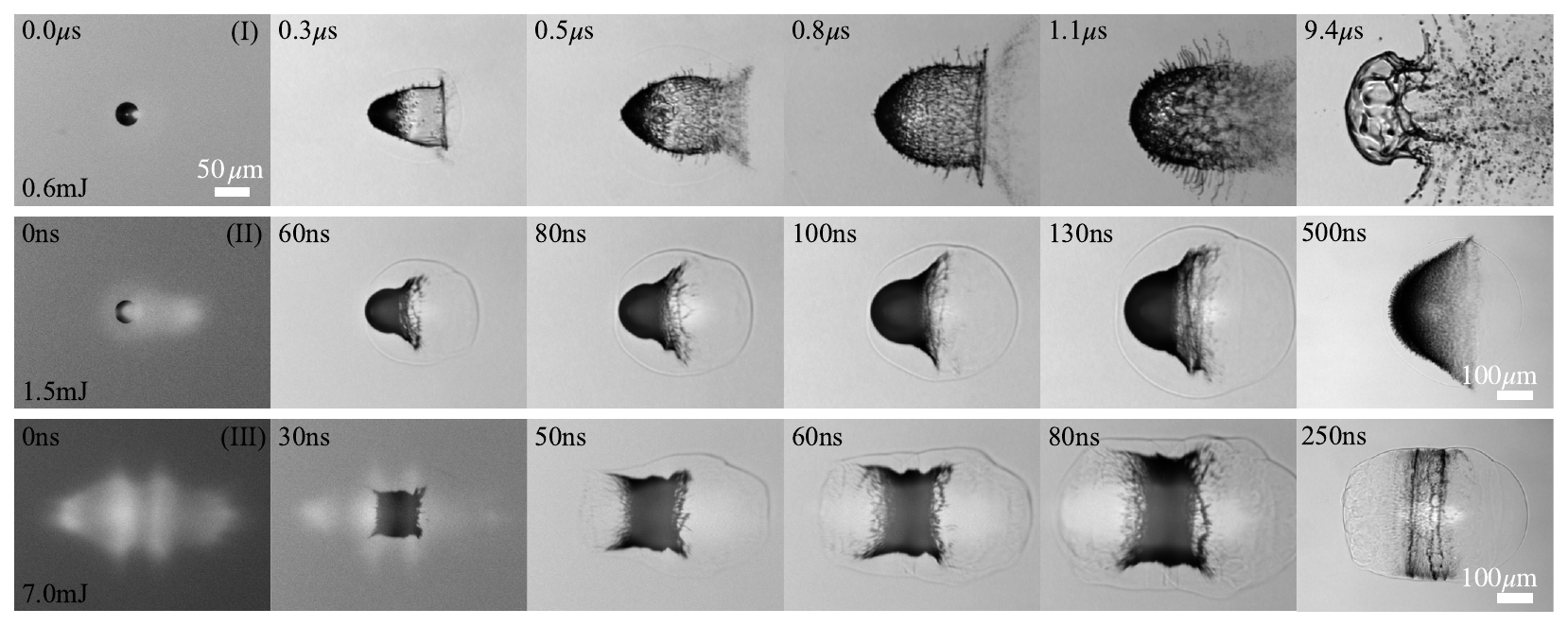}  
    \caption{Droplet deformation modes for different laser energies. At $\el=0.6\,\mathrm{mJ}$, localized plasma formation near the rear interface ($t=0\,\mu\mathrm{s}$) produces an interfacial bubble that expands into a thin liquid sheet. At $\el=1.5\,\mathrm{mJ}$, the plasma extends along the droplet axis and partially into the surrounding air, as observed at $t=0\,\mathrm{ns}$. The droplet subsequently forms an expanding liquid shell that breaks into a dense cloud of fragments by $t=500\,\mathrm{ns}$. At $\el=7.0\,\mathrm{mJ}$, the breakdown region is no longer localized, and the droplet undergoes nearly symmetric breakup. Two shock fronts are emitted from opposite sides of the droplet, while the liquid expands as a cylindrical shell. Both the shock wave and the liquid shell expand at comparable supersonic rates, leading to complete fragmentation by $t=250\,\mathrm{ns}$. The rightmost frames highlight the large difference in timescale required to reach comparable deformation amplitudes at low and high laser energies.}
    \label{fig:phase_diagram}
\end{figure}

We distinguish three main deformation regimes according to the laser energy. Figure\,\ref{fig:phase_diagram} shows time-resolved image sequences for three representative values of $\el$. At low energies, $\el<1\,\mathrm{mJ}$, the deformation is characterized by the formation of an interfacial bubble that subsequently evolves into a thin, expanding liquid film. Closely related cavitation-driven deformation modes have been reported for millimeter-scale droplets and liquid pools with laser-generated bubbles near a free surface \,\cite{avila_cavitation-induced_2015, gonzalez_avila_fragmentation_2016,Hupfeld2020_laser_bubble_interface, rossello2022dynamics,rossello_bubble_2023}. Bubble-driven deformation, splashing, and fragmentation have also been observed in micrometer-sized transparent droplets following optical breakdown \,\cite{chang1988laser_transparent_drop, chylek1987_LIB_drop_gases,zhang1987explosive_vap_larg_drop_laser,Rao2022}. In the millimeter-scale studies, the cavitation location is commonly controlled by externally positioning the laser focus near or within the droplet. In the present experiments, by contrast, the interfacial bubble forms from the optical field established intrinsically by droplet microlensing and internal reflection. Following the initial bubble expansion, capillary forces govern the evolution of the liquid film. The film develops surface corrugations and, depending on laser energy, eventually fragments into radially oriented ligaments. Similar ligament and interfacial structures have been reported in cavitation-driven droplet deformation and in liquid interfaces accelerated by laser-produced bubbles \,\cite{wang2021rayleigh,gonzalez_avila_fragmentation_2016}. In the present images, the ligaments become especially prominent at later times, as illustrated by the frame at $t=1.1\,\mu\mathrm{s}$. Their subsequent stretching and breakup involve capillary mechanisms, including dripping, end pinching, and Rayleigh-Plateau-type destabilization, depending on the local ligament geometry and flow conditions\,\cite{clanet_transition_1999}. Following the initial expansion phase, the droplet evolves into a hollow, bag-like structure at later times, for example at $t=9.4\,\mu\mathrm{s}$, and is propelled at low speeds toward the incoming laser. The axial propulsion velocity $V_\mathrm{ax}$ is substantially smaller than the initial radial expansion rate, $V_\mathrm{ax}\ll\dot{R}_\mathrm{d}$. Close inspection of the rear side of the expanding film also reveals a mist of fine fragments moving with the propagating shock front. The surrounding gas and enveloping shock may generate interfacial shear and contribute to the secondary breakup of these fine structures following Kelvin-Helmholtz instability on top of Rayleigh-Taylor instability, promoting the radial growth of the ligaments. However, distinguishing gas-mediated destabilization from instabilities inherited from the internally accelerated liquid requires the pressure-dependent fragmentation analysis reserved for future work.

For higher laser energies $\el>1.0\,\mathrm{mJ}$, the droplet dynamics change fundamentally. The plasma is no longer confined to a small region near the rear interface but extends along the droplet axis and partially into the surrounding air, as observed at $t=0\,\mathrm{ns}$ for $\el=1.5\,\mathrm{mJ}$ in Fig.\,\ref{fig:phase_diagram}. Axial plasma growth, internal absorption, and plasma emergence beyond transparent-droplet interfaces were already documented in the previous studies with transparent droplets \,\cite{chang1988laser_transparent_drop, chylek1987_LIB_drop_gases,hsieh1987_plasma_front_laser_trans_drop}. The present measurements connect this established plasma behavior to the subsequent transition from connected-film deformation to supersonic liquid-shell fragmentation. The extended plasma produces a stronger and more spatially distributed pressure load within the droplet, resulting in supersonic liquid expansion. The measured shell Mach numbers cover approximately $\mathrm{M}_\mathrm{d}\sim1-12$ in this regime. Although breakup still develops predominantly near the rear side at intermediate energies, the late-time images, for example at $t=500\,\mathrm{ns}$, show a dense cloud of rapidly moving fragments on the laser-facing side. Related transitions from localized breakup to spatially distributed fragmentation have been observed previously when laser energy, breakdown location, or optical configuration was varied \,\cite{Rao2022,Singh_2026}. Here, the transition occurs without changing the external focal position. The increased spatial extent of the ablated region also changes the coupling between the liquid and the emitted shock. This is reflected in the loss of spherical symmetry of the shock front, visible between $60$ and $130\,\mathrm{ns}$ for $\el=1.5\,\mathrm{mJ}$. Although no well-defined cavitation bubble is resolved in this regime, the rear portion of the droplet contains a mixture of fragments, ligaments, and vaporized liquid within the region enclosed by the shock front. Individual fragments become increasingly blurred when their displacement during the $5\,\mathrm{ns}$ illumination pulse exceeds the approximately $5\,\mu\mathrm{m}$ optical resolution, corresponding to velocities above approximately $1\,\mu\mathrm{m/ns}$. 

A further increase in laser energy leads to nearly simultaneous breakup at both the rear and laser-facing sides of the droplet, as observed at $t=30\,\mathrm{ns}$ for $\el=7.0\,\mathrm{mJ}$, as shown in Fig.\,\ref{fig:phase_diagram}. Multidirectional breakup associated with competing internal, interfacial, and near-field breakdown locations has also been reported in transparent millimeter-scale droplets \,\cite{Singh_2026}. In the present geometry, the nearly symmetric response follows the extension of the ablation region along the droplet axis, produced by the combined transmitted and internally reflected optical fields [Fig.\,\ref{fig:optical_breakdown}(c)]. Under these conditions, the droplet expands as an approximately cylindrical liquid shell and becomes completely fragmented by $t=250\,\mathrm{ns}$. Remarkably, the expanding liquid shell and emitted shock attain comparable early-time velocities, indicating strong coupling of both phases to the plasma-generated pressure load. The complete disintegration occurs within a small fraction of the capillary timescale, $t<0.02\tau_c$. 
The sequence of deformation modes resembles regimes previously produced by varying the externally imposed focal position and pulse energy in transparent droplets \,\cite{gonzalez_avila_fragmentation_2016,Singh_2026}. The present measurements provide a quantitative distinction between subsonic deformation modes at $\el<1\,\mathrm{mJ}$ and supersonic shell-fragmentation modes at $\el>1\,\mathrm{mJ}$. The former retain a connected liquid film whose evolution is governed predominantly by inertia and capillarity, whereas the latter produce a fragmented, ballistically expanding liquid shell. Although the individual morphologies have precedents in the literature, their emergence as a continuous energy-controlled sequence under fixed external focusing conditions is the defining feature of the present phase diagram.  

\subsection{Thin film deformation at low energies}\label{subsec:thin_film_deformation}

\begin{figure}
    \centering
    \includegraphics[width=0.9\textwidth]{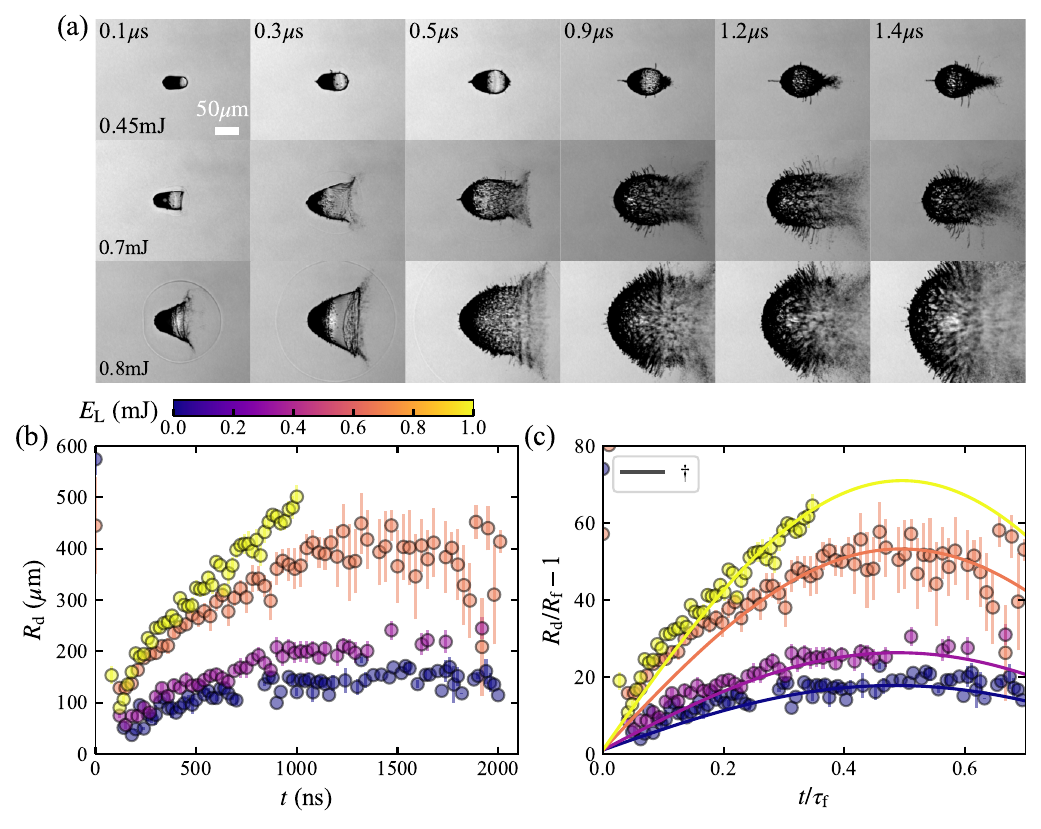}  
    \caption{Evolution of the thin liquid film at low laser energies. (a) Illustrative frames over time for different energies where a thin liquid film is present. Qualitatively, three different liquid films can be distinguished: connected bubble ($\el=0.45\,\mathrm{mJ}$), retracting thin film ($\el=0.7\,\mathrm{mJ}$), and fragmenting thin film ($\el=0.8\,\mathrm{mJ}$; see main text for details). (b) Evolution of the droplet radius $R_\mathrm{d}$ over time for different laser energies at $\lesssim1\,\mathrm{mJ}$. (c) Rescaled droplet radius $R_\mathrm{d}/R_\mathrm{f}$ over dimensionless time $t/\tau_\mathrm{f}$. The solid lines correspond to Eq.\,\eqref{eq:film_equation_solution}.}
    \label{fig:liquid_film}
\end{figure}

Qualitatively, we identify different thin-film behaviors, as depicted in Fig.\,\ref{fig:liquid_film}(a). As the laser energy increases, we first observe a nearly closed bubble that expands and retracts under the competing effects of plasma-driven inertia and capillary forces (see frames at $\el=0.45\,\mathrm{mJ}$). However, the appearance of a long tail on the rear side of the droplet at later times suggests partial opening and collapse of the bubble, as observed at $t=1.4\,\mu\mathrm{s}$. At higher energies, the droplet interface ruptures, giving rise to a liquid film that evolves under a similar mechanical balance, as illustrated for $\el=0.7\,\mathrm{mJ}$. Further increasing the laser energy causes the expansion rate to approach supersonic values, leading to a higher degree of fragmentation, although an intact thin film is still visible during the early stages of the expansion (see frames at $\el=0.8\,\mathrm{mJ}$). In this regime, the number of radially expanding ligaments increases significantly at later times, making it increasingly difficult to distinguish the underlying liquid film. Related bubble-like morphologies have been qualitatively discussed in micrometric-sized transparent droplets\,\cite{wood1988trans_drop_laser_water_ethanol,Rao2022,Singh_2026}. 

The clear expansion and subsequent retraction of the liquid film motivate a quantitative analysis of the observed dynamics. In the thin-film limit, $h(t)\ll R_0$, previously developed models for impulsively accelerated liquid films can be applied, following Refs.\,\cite{vledouts_explosive_2016, Kurilovich2018_cavitation}. Assuming that the film is formed from a fraction of the initial droplet mass characterized by a radius $R_\mathrm{f}\propto R_0$, conservation of mass yields $4/3\pi R_\mathrm{0}^3=4\pi R_\mathrm{d}(t)^2h(t)$, from where thickness over time follows $h(t)=R_\mathrm{0}^3/3R_\mathrm{d}(t)^2$ (assuming negligible mass loss during the explosive expansion). The film dynamics can then be described using the Rayleigh-Plesset equation, where the inertia of the liquid film is balanced by the Laplace pressure $-2\sigma/R_\mathrm{d}(t)$, in the limit of a bubble burst (see Ref.\,\cite{Kurilovich2018_cavitation}). In the thin-film limit, the mechanical balance may be written as $h(t)\ddot{R}_\mathrm{d}\approx\Delta p/\rho$, where $\Delta p=p(R_\mathrm{d}-h)-p(R_\mathrm{d})$ is the pressure difference on the film. Assuming the Laplace pressure as the governing pressure source, it follows: 

\begin{equation}\label{eq:film_equation_differential}
    \ddot{R}_\mathrm{d}(t)=\frac{-2\sigma}{\rho h(t)R_\mathrm{d}(t)}=\frac{-6\sigma R_\mathrm{d}(t)}{\rho R_\mathrm{0}^3}.
\end{equation}

The solution to Eq.\,\eqref{eq:film_equation_differential} was presented in Ref.\,\cite{Kurilovich2018_cavitation} and is: 

\begin{equation}\label{eq:film_equation_solution}
    \frac{R_\mathrm{d}}{R_\mathrm{0}}=\cos(\pi T)+\sqrt{\frac{\We_\mathrm{f}}{12}}\sin(\pi T),
\end{equation}
where $T=t/\tau_\mathrm{f}$ is the dimensionless time, with the film oscillation timescale defined as $\tau_\mathrm{f}=\pi\sqrt{\rho R^3_\mathrm{f}/6\sigma}$.
Equation\,\eqref{eq:film_equation_solution} depends on both $\We_\mathrm{f}$ and $\tau_\mathrm{f}$, which in turn require an estimate of the initial mass involved in bubble formation, characterized by $R_\mathrm{f}$. Inspection of the plasma size in the early-time images of Fig.\,\ref{fig:optical_breakdown}(a) for $\el<1\,\mathrm{mJ}$ suggests that $R_\mathrm{f}\approx0.1-0.2R_0$ is a reasonable estimate. 
Figure\,\ref{fig:liquid_film}(b) shows representative radial expansion curves measured halfway between the rear droplet interface and the rightmost edge of the liquid film. The data exhibit significant scatter due to unavoidable fluctuations in laser-droplet alignment, which introduce variations in the deposited laser energy (see Sec.\,\ref{sec:experimental}). Nevertheless, a reasonable agreement between the experimental observations and Eq.\,\eqref{eq:film_equation_solution} is found, as shown in Fig.\,\ref{fig:liquid_film}(c), where the dimensionless film radius $R_\mathrm{d}/R_\mathrm{f}$ is plotted against the dimensionless time $t/\tau_\mathrm{f}$, where $R_\mathrm{f}$ is kept as a free common fitting parameter, yielding $R_\mathrm{f}\sim0.17R_0$. Thus, the typical values for thin-film expansion rates lie within $100-350\,\mathrm{m/s}$, corresponding to $2200\lesssim\We_\mathrm{f}\lesssim35000$. 
The limitations of this comparison arise not only from the uncertainty in the choice of $R_\mathrm{f}$, but also from the actual behavior of the expanding film. The emergence of radial ligaments and circular ruptures disrupts the ideal free expansion assumed by the model. Additionally, the film increasingly deviates from a spherical geometry as the laser energy increases because of interface breakup and shock wave release into the surrounding air. Despite these limitations, Eq.\,\eqref{eq:film_equation_solution} captures the main qualitative features of the observed liquid-film dynamics. Finally, the thin film regime occupies only a narrow transition region between the optical breakdown threshold and the onset of complete droplet atomization, approximately $0.4\,\mathrm{mJ} \lesssim \el \lesssim 0.8\,\mathrm{mJ}$. Owing to this limited range of accessible conditions, a systematic quantitative scaling between the film kinetic energy and the laser energy cannot be established with sufficient confidence (see Sec.\,\ref{sec:results}\,\ref{subsec:energy_partition}).  

\subsection{Droplet shell and shock wave dynamics at high energies}\label{subsec:droplet_shock_dynamics}

\begin{figure}
    \centering
    \includegraphics[width=1\textwidth]{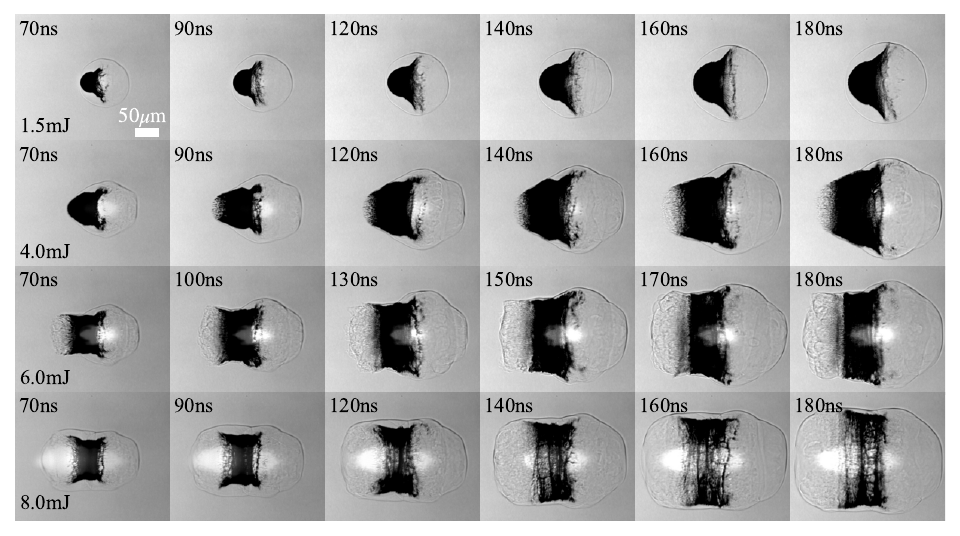}  
    \caption{Evolution over time of the shock wave front as a function of laser energy. For low laser energies (e.g. $\el=1.5\,\mathrm{mJ}$), the strongly localized plasma formed near the rear droplet/air interface leads to a nearly spherical shock front. With increasing laser energy, the laser-facing droplet/air interface breaks due to liquid vaporization ($\el=4.0\,\mathrm{mJ}$, $90\,\mathrm{ns}$ onward), from where a second shock front can emerge ($\el=6.0\,\mathrm{mJ}$). Further increase in laser energy leads to the axial ablation of the droplet resulting in hollowed liquid shell ($\el=8.0\,\mathrm{mJ}$). }
    \label{fig:shock_waves}
\end{figure}

\begin{figure}
    \centering
    \includegraphics[width=0.9\textwidth]{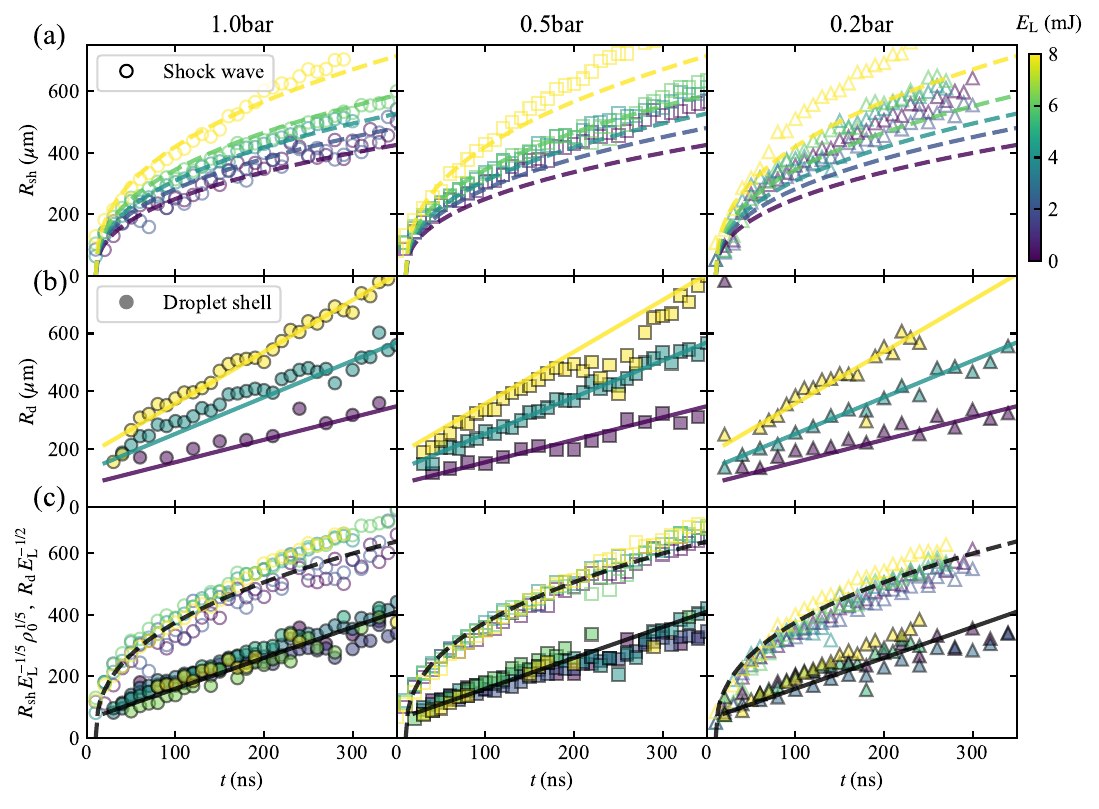}  
    \caption{Expansion dynamics of the droplet shell and the associated shock wave. (a) Shock wave radius $R_\mathrm{sh}$ measured over time on the laser axis for different energies ($0.5-8.0\,\mathrm{mJ}$) at different ambient pressures, $p_0$. The dashed lines correspond to theoretical predictions of a spherical blast wave depicted by Eq.\,\eqref{eq:spherical_blast_R} with $E\sim\eta_\mathrm{s}\el$, where $\eta_\mathrm{s}=0.25\pm0.05$ and $\rho_0\sim1.2\,\mathrm{kg/m^3}$ corresponding to the air density at ambient pressure $p_0=1\,\mathrm{bar}$. (b) Droplet radial extension $R_\mathrm{d}$ over time along the shock front in (a). The solid lines represent the ballistic motion depicted by Eq.\,\eqref{eq:ballistic_motion} with $E\sim\eta_\mathrm{k}\el$, where $\eta_\mathrm{k}=0.45\pm0.07$ (see main text for discussion). (c) Rescaled droplet $R_\mathrm{d}\,\el^{-1/2}$ and shock wave $R_\mathrm{sh}\,\el^{-1/5}\,\rho_0^{1/5}$ radial extensions, allowing the collapse of all available data on the corresponding master curves that illustrate the ballistic motion of the droplet shell, $R_\mathrm{d}\sim t$ [solid line; predicted by Eq.\,\eqref{eq:ballistic_motion}], and nonlinear progression of the shock wave $R_\mathrm{sh}\sim t^{2/5}$ [dashed line; predicted by Eq.\,\eqref{eq:spherical_blast_R}].}
    \label{fig:droplet_and_shock_dynamics}
\end{figure}

A closer examination of the transition from low to high laser energies reveals details regarding the interaction between the shock wave and the droplet, and the resulting deformation dynamics. Figure\,\ref{fig:shock_waves} presents the evolution of both the shock wave and the droplet morphology over the energy range $\el=1.5-8.0\,\mathrm{mJ}$. As the laser energy increases, the ablated region progressively extends toward the laser-facing side of the droplet, creating two distinct source locations for shock wave emission into the surrounding air. Measured from the rupture point, the shock wave appears propagating with an approximately spherical front. According to blast-wave (BW) theory, originally developed in the seminal works of Taylor\,\cite{taylor1950formation}, and later extended in numerous studies\,\cite{sakurai1953blast,Ramsden1964_radiative_blast,sachdev1972propagation}, the evolution of the shock front radius follows the scaling:
\begin{equation}\label{eq:blast_wave_general}
    R(t)=\xi(\gamma)\left(\frac{E}{\rho_0}\right)^{1/(n+2)}t^{2/(n+2)},
\end{equation}
where $E$ is the energy deposited into the blast wave, $\rho_0$ is the ambient density, and $\xi(\gamma)\simeq1$ is a dimensionless constant that depends on the adiabatic index $\gamma\simeq1.4$, given under standard conditions. The exponent $n$ reflects the dimensionality of the energy release: $n=1$ for planar, $n=2$ for cylindrical, and $n=3$ for spherical expansions\,\cite{girin2022compressible_dynamics}. Considering the optical breakdown as a nearly point-like energy release on the length scale of the observed shock wave propagation, Eq.\,\eqref{eq:blast_wave_general} yields: 

\begin{subequations}\label{eq:spherical_blast}
\begin{align}
    R_\mathrm{sh}(t) &\approx \left(\frac{E}{\rho_0}\right)^{1/5} t^{2/5}, \label{eq:spherical_blast_R}\\
    V_\mathrm{sh}(t) &= \frac{dR}{dt} = \frac{2}{5}\left(\frac{E}{\rho_0}\right)^{1/5} t^{-3/5}. \label{eq:spherical_blast_V}
\end{align}
\end{subequations}

From Eq.\,\eqref{eq:spherical_blast_R}, the shock-front evolution depends on both the energy deposited into the blast wave and the density of the propagation medium. From Fig.\,\ref{fig:shock_waves} we observe at high laser energies a strong interaction between the shock and liquid shell. For consistency, the shock front is measured from the rightmost edge. In Fig.\,\ref{fig:droplet_and_shock_dynamics}(a), we illustrate this dependence by showing the shock-front radius $R_\mathrm{sh}$ as a function of time for different laser energies and ambient pressures ranging from $0.1$ to $1.0\,\mathrm{bar}$. Assuming an air density of $\rho_0\simeq1.2\,\mathrm{kg/m^3}$ at $1\,\mathrm{bar}$, the experimental data are well described by the blast-wave scaling given by Eq.\,\eqref{eq:spherical_blast_R}. The nonlinear fits in Fig.\,\ref{fig:droplet_and_shock_dynamics}(a) further allow us to estimate the energy deposited into the blast wave as a fraction of the laser energy, $E\simeq\eta_\mathrm{s} E_\mathrm{L}$, where $\eta_\mathrm{s}$ remains as a free fitting parameter. At sufficiently high laser energies, we find $\eta_\mathrm{s}=0.25\pm0.05$, indicating that approximately $25\%$ of the incident laser energy is transferred to the emitted shock wave. Thus, the power-law depicted by Eqs.\,\eqref{eq:spherical_blast} match well with the experimental data. Previous studies of nanosecond laser-induced optical breakdown in bulk water have reported energy fractions exceeding approximately $40\%$\,\cite{Vogel_LIP_thresholds,Vogel_1999,Linz_2025}. The lower fraction observed in the present experiments may result from the confined droplet geometry and the increasing transfer of energy to the liquid as the plasma extends along the droplet axis at higher laser energies\,\cite{Vogel_LIP_thresholds,Vogel_1999,Linz_2025}. Studies of micrometer-sized water droplets have reported shock wave energy fractions comparable to those measured here, with an average value of approximately $30\%$ of the incident laser energy \,\cite{Rao2022}.

The role of the ambient gas density on the shock is further highlighted by comparing shock-wave expansion curves obtained at similar laser energies but different ambient pressures. We plot the rescaled shock radius $R_\mathrm{sh}E_\mathrm{L}^{-1/5}\rho_0^{1/5}$ as a function of time for all laser energies and ambient pressures. Here, $\rho_0$ is estimated from the ideal-gas relation using the measured ambient pressure and room temperature. As shown in Fig.\,\ref{fig:droplet_and_shock_dynamics}(c), this rescaling collapses the shock wave data onto a single master curve. Interestingly, while the shock wave expansion exhibits the expected dependence on ambient density, the droplet-shell dynamics remain largely insensitive to the surrounding pressure. At fixed laser energy, the nearly identical shell trajectories indicate that the effective mechanical energy and impulse coupled to the liquid do not change detectably over the investigated pressure range. This statement concerns the mechanically deposited energy inferred from the shell motion, rather than the total absorbed laser energy, which is not measured directly. This observation is consistent with earlier measurements showing that the optical-breakdown threshold of water droplets in air can remain nearly independent of ambient pressure when breakdown is initiated inside the liquid\,\cite{chylek1990pressure}. Whereas those studies focused on the breakdown threshold and initiation location, the present measurements show that the subsequent mechanical loading and deformation of the droplet are also nearly pressure-independent. By contrast, the shock-wave trajectory changes with the density of the surrounding gas, as expected from blast wave scaling. The different pressure sensitivities of the liquid shell and external shock support the interpretation that both originate from a common internal plasma-generated pressure load. The surrounding gas controls the propagation of the shock after its release into air, but it does not govern the deformation of the droplet itself.

The droplet deformation can be rationalized by assuming that a constant fraction of the incident laser energy is converted into kinetic energy of the expanding liquid shell. We heuristically consider that droplet shell kinetic energy $E_\mathrm{k}=\frac{1}{2}M_0\left(dR_\mathrm{d}/dt\right)^2$ scales linearly with the laser energy, $E_\mathrm{k}=\eta_\mathrm{k}\el$. Rewriting the latter equation leads to $dR_\mathrm{d}/dt\approx(2\eta_\mathrm{k}\el/M_0)^{1/2}$, that after integration results in:

\begin{equation}\label{eq:ballistic_motion}
    R_\mathrm{d}\approx\left(\frac{2\eta_\mathrm{k}\el}{M_0}\right)^{1/2}\,t+R_0.
\end{equation}

\begin{figure}
    \centering
    \includegraphics[width=1.0\textwidth]{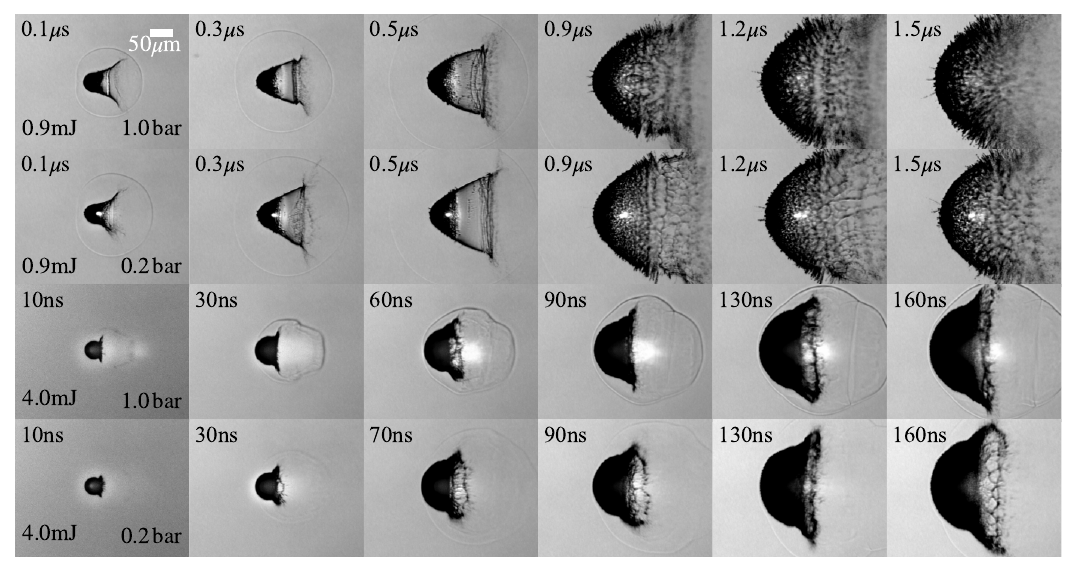}  
    \caption{Evolution of the droplet over time at constant laser energy and different ambient pressures. Two cases are shown: thin film regime at low laser energies $\el=0.9\,\mathrm{mJ}$, and ballistic liquid shell regime observed at higher laser energies $\el=4.0\,\mathrm{mJ}$. Both energies are compared at two different pressures $0.2$ and $1.0\,\mathrm{bar}$. Both regimes are illustrated within the characteristic timescale of the droplet evolution, being $\sim1-10\,\mu\mathrm{s}$ and $\sim100-300\,\mathrm{ns}$ for film and shell, respectively. }
    \label{fig:droplet_comparison_pressure}
\end{figure}

\begin{figure}
    \centering
    \includegraphics[width=0.6\textwidth]{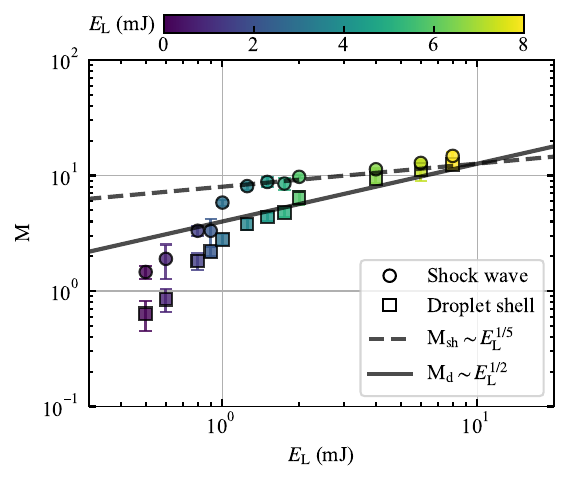}  
    \caption{Comparison of the dimensionless expansion rate $\mathrm{M}=\dot{R}/c$ as a function of laser energy $\el$ for shock waves (circles) and droplet shell (squares). The dashed line corresponds to the scaling dependence $\mathrm{M}_\mathrm{sh}\sim\el^{1/5}$ as deduced from BW theory for the shock wave front. The solid line corresponds to the scaling dependence $\mathrm{M}_\mathrm{d}\sim\el^{1/2}$ as inferred from energy conservation arguments for droplet expansion dynamics. Both lines are power-law fits, with the constant of proportionality as the sole fitting parameter.}
    \label{fig:expansion_rates}
\end{figure}

Equation\,\eqref{eq:ballistic_motion} predicts ballistic expansion of the liquid shell, with a characteristic velocity that scales with laser energy as $\dot{R}_\mathrm{d}\sim\el^{1/2}$. Note that similar dependence has been observed in the framework of tin droplet propulsion after laser impact in full ablation regime\,\cite{kurilovich2018power}. In Fig.\,\ref{fig:droplet_and_shock_dynamics}(b), we show the droplet expansion curves for different laser energies and ambient pressures. Linear fits to the data yield an estimated upper energy fraction of approximately $\eta_\mathrm{k}=0.45\pm0.07$ for all datasets. Thus, the estimated kinetic energy of the expanding liquid shell is approximately twice the energy carried by the shock wave, for which the inferred fraction approaches $\eta_\mathrm{s}=0.25\pm0.05$ at sufficiently high laser energies. Importantly, Eq.\,\eqref{eq:ballistic_motion} contains no explicit dependence on the ambient pressure. This pressure independence is confirmed by the measurements in Fig.\,\ref{fig:droplet_and_shock_dynamics}(b), where the droplet expansion curves remain unchanged over the explored pressure range. When the shell radius is rescaled as $R_\mathrm{d}\el^{-1/2}$, the experimental data collapse onto a single master curve, as shown in Fig.\,\ref{fig:droplet_and_shock_dynamics}(c). This collapse is consistent with ballistic expansion of the liquid shell. This result is further supported by a direct comparison of droplet morphologies at fixed laser energy and different ambient pressures. Figure\,\ref{fig:droplet_comparison_pressure} shows two representative examples: one in the thin-film regime ($\el=0.9\,\mathrm{mJ}$) and one in the shell-expansion regime ($\el=4.0\,\mathrm{mJ}$), each recorded at $0.2$ and $1.0\,\mathrm{bar}$. In both regimes, the bulk dynamics of the droplet remain nearly identical. This behavior is consistent with the inertial pressure scale associated with the expanding liquid, $\rho_\mathrm{w}\dot{R}_\mathrm{d}^{2}$, greatly exceeding the ambient pressure $p_0$. Even for the lowest measured film expansion velocity, $\dot{R}_\mathrm{d}\sim150\,\mathrm{m/s}$, this pressure scale is approximately $22\,\mathrm{MPa}$, more than two orders of magnitude larger than atmospheric pressure. The direct influence of ambient pressure on the liquid expansion is therefore expected to be negligible. More importantly, reducing the ambient pressure substantially modifies the density-dependent propagation of the external shock wave and the conditions for gas-phase breakdown, while leaving the droplet morphology and expansion rate unchanged (see Fig.\,\ref{fig:droplet_and_shock_dynamics} and Fig.\,\ref{fig:droplet_comparison_pressure}). This observation rules out external air plasma and air-mediated loading as the primary drivers of droplet deformation, supporting instead an impulse generated by plasma formation within the liquid. Finally, several interfacial instabilities exhibit qualitatively different behavior at different ambient pressures. A systematic investigation of these pressure-dependent instabilities will be reported in future work.

Figure\,\ref{fig:shock_waves} shows that the shock front and liquid shell can exhibit comparable early-time expansion velocities, particularly at the highest laser energies. Unlike the approximately constant shell velocity in the ballistic regime, the shock velocity decreases as $V_\mathrm{sh}\sim t^{-3/5}$ within the spherical Taylor-Sedov approximation. A meaningful comparison across laser energies must therefore be performed at a common reference time, as shown in Fig.\,\ref{fig:expansion_rates}. Here, both the shock and shell velocities are evaluated from $t=0$ to $t_\mathrm{ref}=50\,\mathrm{ns}$ by differentiating the fitted trajectories. The corresponding dimensionless velocity of the shock front $\mathrm{M}_\mathrm{sh}=V_\mathrm{sh}/c$ follows the predicted scaling $V_\mathrm{sh}\sim\el^{1/5}$ at sufficiently high laser energies. A pronounced deviation appears below $\mathrm{M}_\mathrm{sh}\approx4$, marking the transition toward the weak-shock regime\,\cite{Ramsden1964_radiative_blast, Wei_2021_weak_strong_BW}. In this regime, the ambient pressure can no longer be neglected and the strong, self-similar Taylor-Sedov solution progressively loses validity.

The liquid-shell expansion follows a different scaling. Assuming that a constant fraction of the laser energy is converted into kinetic energy of the shell, Eq.\,\eqref{eq:ballistic_motion} predicts $\dot{R}_\mathrm{d}\sim\el^{1/2}$ or, equivalently, $\mathrm{M}_\mathrm{d}=\dot{R}_\mathrm{d}/c\,\sim \el^{1/2}$. As shown in Fig.\,\ref{fig:expansion_rates}, the measured shell expansion rates agree well with this prediction at sufficiently high laser energies, where the approximately constant energy-partition ratio $E_\mathrm{k}/\el\approx0.45$ remains valid. At lower laser energies, significant deviations from the ballistic scaling are observed. As discussed in Fig.\,\ref{fig:phase_diagram}, the ballistic expansion regime is progressively replaced by a more complex thin-film deformation for $\el<1\,\mathrm{mJ}$, where inertial and capillary effects jointly govern the droplet dynamics.

\subsection{Energy partition}\label{subsec:energy_partition}

\begin{figure}
    \centering
    \includegraphics[width=0.8\textwidth]{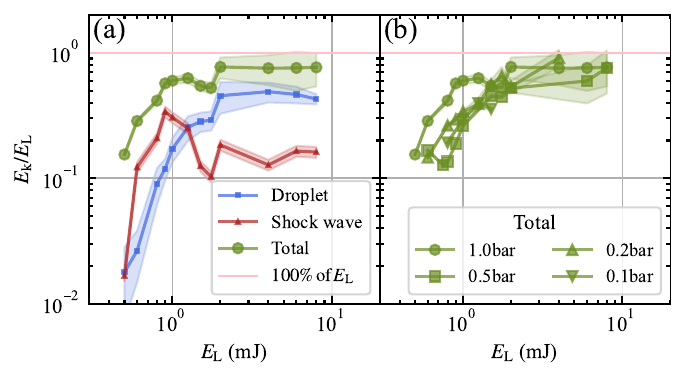}  
    \caption{(a) Fraction of the energy $E_\mathrm{k}/\el$ estimated from the shell kinetic energy (based on radial expansion rate) and energy associated with the corresponding shock wave. The sum of these two contributions yields $\sim70\,\%$ of the laser energy. (b) Sum of the shell and shock energies at different surrounding air pressures.}
    \label{fig:Energy_partition}
\end{figure}

Finally, we consider the partitioning of the incident laser energy between the mechanically resolved channels. The shock wave and liquid shell energies are inferred from the expansion dynamics presented in Fig.\,\ref{fig:droplet_and_shock_dynamics}. The liquid kinetic energy is estimated under the assumption that the entire initial droplet mass moves with a single characteristic expansion velocity, which is inconsistent with $R_\mathrm{f}\sim0.17R_0$ at low energies. This approximation neglects the spatial distributions of liquid mass and velocity within the expanding structure and should therefore be interpreted as an upper bound on the liquid kinetic energy. The approximation is expected to become more representative at $\el\gtrsim4\,\mathrm{mJ}$, where the observed liquid contour expands more uniformly and approaches a cylindrical morphology.

The transition from thin film formation to liquid shell expansion occurs near $\el\sim1\,\mathrm{mJ}$. The fraction of energy carried by the shock wave reaches a maximum near this transition and subsequently decreases, whereas the estimated liquid kinetic energy fraction progressively approaches a plateau. For $\el<1\,\mathrm{mJ}$, the energy carried by the shock wave exceeds the estimated kinetic energy of the thin film. This behavior is consistent with plasma localization near the droplet-air interface, which enables rapid release of the pressure load into the surrounding air following interfacial rupture. At higher energies, the plasma and ablated region extend along the droplet axis [Fig.\,\ref{fig:optical_breakdown}], increasing the mechanical coupling between the plasma and the liquid. The plateau values of the estimated shock wave and liquid kinetic energies approach approximately $0.25\el$ and $0.45\el$, respectively.

The two dominant mechanically resolved channels therefore account for an upper value of approximately $70\%$ of the incident laser energy, as shown in Fig.\,\ref{fig:Energy_partition}. A comparable overall energy fraction has been reported previously for laser-induced droplet breakup\,\cite{Rao2022}. In that study, however, the breakup energy was not determined directly from the fragment mass and velocity distributions. Instead, it was inferred as the residual energy remaining after subtracting the measured transmitted energy and the model-derived shock wave energy from the incident pulse energy. Consequently, the breakup energy fraction reported in Ref.\,\cite{Rao2022} may also include unresolved thermal, optical, and other non-shock energy channels. For nanosecond laser-induced optical breakdown in bulk water, approximately $70-85\%$ of the incident laser energy has been estimated to contribute to mechanical effects\,\cite{Vogel_1999}. Approximately $50\%$ of the incident energy may be stored in the cavitation bubble, depending on the irradiation conditions \,\cite{Vogel_1999}. In the present confined geometry, the nearby free surface ruptures before a cavitation bubble can expand to a well-defined maximum volume. The liquid response therefore differs fundamentally from that in bulk water, with a substantial fraction of the mechanically coupled energy appearing instead as deformation and fragmentation of the droplet. A smaller mechanical contribution arises from the axial motion of the droplet. At low energies, asymmetric plasma localization near the rear side produces weak propulsion toward the incoming laser, barely visible in Fig.\,\ref{fig:phase_diagram}. Because the axial velocity is much smaller than the radial expansion velocity, $V_\mathrm{ax}\ll\dot{R}_\mathrm{d}$, the corresponding kinetic energy accounts for at most approximately $1\%$ of the incident laser energy.

In the absence of direct optical-transmission and calorimetric measurements, these contributions cannot be separated quantitatively. For reference, heating the entire initial droplet from $T_0=20^\circ\mathrm{C}$ to the normal boiling temperature $T_\mathrm{f}=100^\circ\mathrm{C}$ requires $E_\mathrm{heat} =M_0c_p(T_f-T_0) \approx0.13\,\mathrm{mJ}$,  where $c_\mathrm{p}\simeq4180\,\mathrm{J\,kg^{-1}K^{-1}}$. Complete vaporization would require an additional $E_\mathrm{vap} =M_0\Delta H \approx0.86\,\mathrm{mJ}$ where $\Delta H\simeq2.26\times10^6\,\mathrm{J/kg}$. Heating and completely vaporizing the initial droplet would therefore require approximately $1.0\,\mathrm{mJ}$. These values represent characteristic energy scales rather than measured energy fractions because only an unknown fraction of the liquid is heated or vaporized. Previous measurements of nanosecond optical breakdown in bulk water attributed approximately $7.5\%$ of the incident energy to vaporization \,\cite{Vogel_1999}; however, this value cannot be transferred directly to the present freely deforming droplet geometry.  
In addition to liquid kinetic energy and thermal energy fraction, 
for a centered Gaussian beam, approximately $6\%$ of the incident energy geometrically bypasses the droplet. Accounting for the measured laser-droplet alignment variation increases this estimate to at most approximately $8\%$. After including this geometric bypass and the estimated mechanical channels, approximately $20\%$ of the incident energy remains unresolved. This remainder may be distributed among transmitted, reflected, and scattered light, plasma radiation, liquid heating, partial vaporization, and unresolved liquid motion. In bulk water, plasma radiation and reflected or scattered light have been reported to contribute only small fractions of the incident energy, $<2\%$, whereas approximately $8\%$ was transmitted and approximately $10\%$ remained unassigned\,\cite{Vogel_1999}. These values provide useful reference scales but cannot be applied quantitatively to the present droplet experiments. Because the transmitted optical energy and thermal channels are not measured here, the unresolved fraction cannot be partitioned further.

\section{Conclusions}

In this study, we presented a quantitative characterization of the droplet and shock wave dynamics following nanosecond laser-induced optical breakdown in micrometer-sized water droplets. For comparable beam and droplet diameters, the microlensing action of the curved liquid interface produces pronounced internal and near-field intensity maxima. Under the fixed focusing conditions used here, internal reflection enhances the optical intensity near the rear side of the droplet, resulting in plasma formation inside the liquid. The subsequent plasma expansion generates an internal pressure load that drives the inside-out deformation and fragmentation of the droplet. Distinct deformation regimes were identified as a function of laser pulse energy. At low energies, the plasma remains localized near the rear side of the droplet and produces an interfacial bubble. The bubble expands into a thin liquid film that subsequently retracts under the action of surface tension, while developing surface corrugations and radially oriented ligaments. Part of the plasma-generated pressure is simultaneously released into the surrounding air as a nearly spherical shock wave. At higher laser energies, the plasma extends along the droplet axis, producing a more spatially distributed pressure load, multiple shock fronts, and increasingly symmetric breakup. In this regime, the liquid undergoes supersonic shell expansion and ultimately complete fragmentation. 

Although related deformation morphologies have been reported previously, the present experiments provide a quantitative comparison of the liquid-shell and shock-front dynamics under fixed external focusing conditions and changing ambient pressures. Varying only the pulse energy produces the complete transition from interfacial-bubble and thin-film dynamics to supersonic shell expansion and complete fragmentation. Simultaneous measurements of the liquid and gas phases show that the shock front and liquid shell can attain comparable early-time velocities at high laser energies, but subsequently evolve according to distinct scaling laws. Within the strong-shock and approximately spherical regime, the external shock follows a density-dependent Taylor-Sedov scaling. The liquid shell, by contrast, undergoes ballistic expansion, with a characteristic velocity derived based on energy conservation arguments. Rescaling the shock and liquid shell trajectories according to their respective energy and density dependencies collapses measurements obtained over the explored range of pulse energies and ambient pressures. This collapse provides a quantitative separation between the blast wave propagation in the gas and the ballistic expansion of the liquid. 

A central contribution of this work is the use of ambient pressure as a dynamical control parameter. Earlier studies established that ambient conditions can influence the threshold and initiation location of optical breakdown in transparent droplets. The present measurements instead resolve the pressure dependence of the subsequent shock and liquid dynamics. Reducing the ambient pressure changes the propagation of the external shock through its dependence on gas density, while the liquid-shell morphology and expansion rate remain unchanged. The distinct pressure sensitivities of the two phases show that the surrounding gas governs shock propagation after the pressure wave is released from the droplet, but does not provide the dominant impulse responsible for liquid deformation. The pressure independence of the shell kinematics also argues against external air plasma or gas-mediated loading as the principal deformation mechanism. Instead, the results support an internally driven fragmentation process. The estimated shock-wave and liquid kinetic energies approach approximately $0.25E_\mathrm{L}$ and $0.45E_\mathrm{L}$, respectively. The two dominant mechanically resolved channels therefore account for an upper value of approximately $70\%$ of the incident laser energy. Therefore, the result demonstrates substantial coupling of the incident optical energy to mechanically resolved motion in the gas and liquid phases. At lower laser energies, deviations from the blast wave and ballistic shell descriptions become increasingly important. Localized plasma formation produces asymmetric loading, while inertia, capillarity, and surface instabilities govern the expansion and retraction of the connected liquid film. In parallel, the external shock progressively enters the weak shock regime below $\mathrm{M}_\mathrm{sh}\approx4$, where ambient pressure can no longer be neglected and the strong, self-similar blast wave approximation loses validity. Future work should examine whether ambient pressure also influences the later fragmentation of the low-energy liquid film, including surface corrugation, radial-ligament growth, secondary breakup, and the resulting fragment-size distribution. Additional measurements of transmitted optical energy, plasma emission, liquid mass loss, and fragment velocities would further resolve the remaining energy channels and provide a more complete description of pressure-dependent fragmentation following optical breakdown.

\section*{Acknowledgments}

The Advanced Research Center for Nanolithography (ARCNL) is a public-private partnership between the University of Amsterdam (UvA), Vrije Universiteit Amsterdam (VU), Rijksuniversiteit Groningen (UG), the Dutch Research Council (NWO), and the semiconductor equipment manufacturer ASML. This project was partly financed by "Toeslag voor Topconsortia voor Kennis en Innovatie (TKI)" from the Dutch Ministry of Economic Affairs and Climate Policy, the project "Plasma driven by a variable-wavelength laser for next-generation EUV sources for nanolithography" (with Project No. 19458) of the Open Technology Programme (which is financed by NWO), and the European Research Council (CoG Project No. 101086839). The authors thank Fleur Veen and the other members of the Ion Interactions group at the University of Groningen for their assistance with the droplet-generator setup.

\section*{Data availability}
The data that support the findings of this article are not publicly available because they contain commercially sensitive information. The data are available from the authors upon reasonable request.

\newpage

\appendix\label{appendix}

\section{Laser beam modeling with ray tracer}\label{app:ray-tracer}

\begin{figure}
    \centering
    \includegraphics[width=0.7\linewidth]{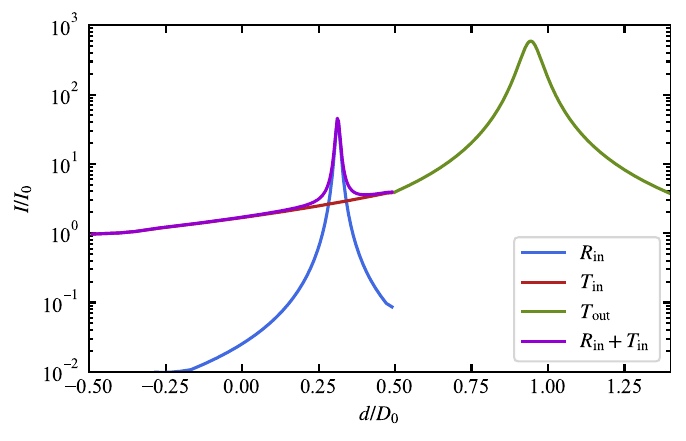}
    \caption{Enhancement curves computed from two-dimensional ray-tracing based on geometric optics, where normalized intensity, with $I_0=1$, is plotted against dimensionless axial coordinate $d/D_0$.}
    \label{app:figure_enhancement_curves}
\end{figure}

In this study, we employ a two-dimensional ray-tracing model based on geometrical optics. The incident laser beam is represented by a set of discrete rays carrying Gaussian-distributed energy:

\begin{equation} 
    w_i=\exp\left(-2\frac{y_i^2}{w_0^2}\right), 
\end{equation}
where $w_0$ is the $1/e^2$ radius of the incident beam and $y_i$ is the transverse position of each ray. The ray energies are subsequently normalized so that the total beam energy equals unity.
At a given axial location $x$, the transverse positions of all rays crossing that plane are determined. The weighted beam centroid is then defined as:

\begin{equation} 
    \hat{y}(x)= \frac{\sum_i w_i y_i(x)} {\sum_i w_i}. 
\end{equation}

The weighted root-mean-square (RMS) quantity is:

\begin{equation} 
    \sigma(x)=\sqrt{\frac{\sum_i w_i\left[y_i(x)-\hat{y}(x)\right]^2}{\sum_i w_i}},
\end{equation}
and the corresponding equivalent Gaussian $1/e^2$ radius is:

\begin{equation}
    w(x)=2\sigma(x).
\end{equation}

The RMS quantity and radii of the incident beam are computed using the same definition:

\begin{equation} 
    \sigma_{\mathrm{inc}}= \sqrt{ \frac{\sum_i w_i\left(y_i-\hat{y}_0\right)^2} {\sum_i w_i} },\,\,\, w_\mathrm{inc}=2\sigma_\mathrm{inc}.
\end{equation}
where $\hat{y}_0$ denotes the centroid of the incident beam. Assuming axisymmetric focusing, the optical intensity scales inversely with the beam cross-sectional area:

\begin{equation} 
    I(x)\propto\frac{1}{w(x)^2}. 
\end{equation}

The local intensity enhancement relative to the incident beam is therefore estimated as:

\begin{equation} 
    \frac{I(x)}{I_0}=\left( \frac{w_{\mathrm{inc}}} {w(x)} \right)^2. 
\end{equation}

This expression provides a geometrical estimate of the optical energy concentration induced by droplet focusing. Additionally, a Fresnel correction factor is computed at each interface, given the angular dependence of the reflection and transmission coefficients. Denoting incident $\theta_\mathrm{i}$ and transmitted angles $\theta_\mathrm{t}$, the power reflectances for the $s$- and $p$-polarized components are calculated:

\begin{equation}
    R_\mathrm{s}=\left|\frac{n_1\cos{\theta_\mathrm{i}-n_2\cos{\theta_\mathrm{t}}}}{n_1\cos{\theta_\mathrm{i}+n_2\cos{\theta_\mathrm{t}}}}\right|^2, \,\,\,
    R_\mathrm{p}=\left|\frac{n_2\cos{\theta_\mathrm{i}-n_1\cos{\theta_\mathrm{t}}}}{n_2\cos{\theta_\mathrm{i}+n_1\cos{\theta_\mathrm{t}}}}\right|^2.
\end{equation}

For circularly polarized light, the effective reflection and transmission coefficients are:

\begin{equation}
    R=\frac{R_\mathrm{s}+R_\mathrm{p}}{2}, \,\,\, T=1-R.
\end{equation}

For propagation from water to air, total internal reflection occurs when $\sin^2\theta_\mathrm{i}\ge1$, or equivalently $\theta_\mathrm{i}\ge\theta_\mathrm{c}$, where the critical angle is:

\begin{equation}
    \theta_\mathrm{c}=\arcsin{\left(\frac{n_\mathrm{air}}{n_\mathrm{w}}\right)}.
\end{equation}

Considering $n_\mathrm{w}=1.33$ and $n_\mathrm{air}=1$, yields $\theta_\mathrm{c}\simeq48.8^\mathrm{o}$. Based on the above formulation, intensity-enhancement curves can be computed for a droplet-to-beam diameter ratio of $\alpha=D_0/D_\mathrm{b}$, as shown in Fig.\,\ref{app:figure_enhancement_curves}. Four different contributions are displayed: rays transmitted from air into the droplet ($T_\mathrm{in}$), internally reflected rays ($R_\mathrm{in}$), the sum of the transmitted and reflected contributions ($R_\mathrm{in}+T_\mathrm{in}$), and rays transmitted from the droplet back into air ($T_\mathrm{out}$). 
The intensity enhancement inside the droplet is dominated by a sharp peak located at approximately $d/D_0=0.35$, corresponding to the region of strongest geometrical focusing. However, the extended tail for $d/D_0<0.35$ remains important, particularly at higher laser pulse energies, as it contributes to a significant increase in the local optical intensity over a larger fraction of the droplet volume. The model provides a gain factor for the reflection branch of $G_\mathrm{r}\approx45$; for the reported minimum peak energy for plasma formation $\el\approx0.5\,\mathrm{mJ}$, the corresponding incident intensity $I_0\sim3.6\times10^9\,\mathrm{W/cm^2}$ increases up to $\approx1.6\times10^{11}\,\mathrm{W/cm^2}$, that is of the same order as the previously reported bulk water optical breakdown under nanosecond laser irradiation $\approx10^{11}\,\mathrm{W/cm^2}$\,\cite{Vogel_LIP_thresholds}. 
Finally, the calculated intensity profiles are regularized by accounting for the diffraction limit, as explained in the main text (see Sec.\,\ref{sec:results}\,\ref{subsec:optical_breakdown}).

\newpage

%


\end{document}